\documentstyle[11pt]{article}
\newcommand{\hs}{\hspace{0.6cm}}
\newcommand{\st}{space-time}
\newcommand{\ie}{{\em i.e.}}
\newcommand{\eg}{{\em e.g.}}
\newcommand{\et}{{\em et al.}}
\newcommand{\th}{theory}

\newenvironment{proof}{{\bf Proof:} \begin{quotation}}{\end{quotation}}
\newenvironment{remark}{{\bf Remark:}  \begin{quotation}}{\end{quotation}}

\newtheorem{cor}{Cor.}
\newtheorem{thm}{T}
\newtheorem{axm}{A}
\newtheorem{dfn}{D}
\title{Steps towards an axiomatic pregeometry of space-time}

\author{S. E. Perez Bergliaffa$^1$\and G. E. Romero$^2$ \and H. Vucetich$^3$}

\begin{document}
\maketitle

\noindent $^1$ Departamento de F\'{\i}sica, Universidad Nacional de La Plata, 
CC 67, CP 1900 La Plata, Argentina\\
$^2$ Instituto Astron\^omico e Geof\'{\i}sico, Universidade de S\~ao Paulo, Av. 
M. 
Stefano 4200, CEP 04301-904, S\~ao Paulo SP, Brazil\\
$^3$ FCAyG, Observatorio de La Plata, Paseo del Bosque s/n, CP 1900 La Plata, 
Argentina\\

\begin{abstract}
We present a deductive theory of space-time which is realistic,
objective, and relational. It is realistic because it assumes the
existence of physical things endowed with concrete properties. It is
objective because it can be formulated without any reference to
cognoscent subjects or sensorial fields. Finally, it is relational
because it assumes that \st\ is not a thing but a complex of relations
among things. In this way, the original program
of Leibniz is consummated, in the sense that space is ultimately an order 
of
coexistents, and time is an order of succesives. In this context,
we show that the
metric and topological properties of Minkowskian \st\ are reduced to
relational properties of concrete things. We also sketch how our theory can 
be extended to encompass a Riemannian \st .
\end{abstract}

\section{Introduction}

\hs Space-time is a primitive (\ie\ non-derivable) concept in every
physical theory.  Even the so-called space-time theories, like General
Relativity, do not deal with the nature of \st\ but with its
geometrical structure. The question ``what is \st ?'' precedes
the formulation of any specific physical theory, and belongs to what
is usually known as Protophysics (\ie\ the branch of scientific ontology
concerned with the basic assumptions of Physics).\\ 

We must recall that the ontological status of space-time has
been a subject of debate for physicists and philosophers during the
last 300 years. The kernel of this debate has been the confrontation
of two antagonic positions: absolutism and relationalism. The former
considers \st\ as much a thing as planets and electrons are,
\ie\ \st\ would be a physical entity endowed with concrete
properties. This is the position held by Newton in his renowned
discussion with Leibniz (mediated by S. Clarke: Alexander, 1983), and also
by J. Wheeler in the geometrodynamical approach to physics (\eg\ Misner
\et, 1973). The relationalism instead asserts that \st\ is not a
thing but a complex of relations among physical things. In Leibniz's
words: ``I have said more than once, that I hold space to be something
merely relative, as time is; that I hold it to be an order of
coexistents, as time is an order of successions'' (see Alexander
1983).\\

An important consequence of Leibniz's ideas is that if \st\ is not an
ontological primitive, then it should be possible to construct it
starting from a deeper ontological level. That is to say, the
spatiotemporal relations should be definable from more fundamental
relations. There have been several attempts to demonstrate the
relational nature of \st, both subjectivistic and phenomelogical
(\eg\ Carnap 1928 and Basri 1966) and objective and realistic (Bunge
and Garc\'{\i}a Maynez 1977, and Bunge 1977). We think that a
deductive theory of \st\ cannot be built with blocks that are
alien to the physical discourse (such as cognoscent subjects or
sensorial fields) in order to be compatible with contemporary physical
theories. In this sense, we coincide with Bunge's approach, which only
assumes the presuppositions common to the entire physical science (see
Bunge 1977).\\

We present here a new formulation, realistic and objective, of the
relational theory of \st, based on the scientific ontology
of Bunge (1977, 1979).  The \th\ will be displayed as an axiomatic
system, in such a way that its structure will turn out to be easily
analyzable \footnote{On the advantadges of the axiomatic method see Perez
Bergliaffa \et\ (1993) and references therein.}. The construction of the theory 
lays on the notion of interaction among basic things, and on the notion
of simultaneity.\\

At this point, we should mention that the search for a Quantum Theory of
Gravity has triggered an intense research on the nature of \st\ (for an 
exhaustive review, see Gibbs 1996). The aim of this research is to build a 
theory (``pregeometry'') from which all the ``properties'' of \st\ 
(like continuity and dimensionality) can be explained \footnote{Let us 
recall that,
according to our view, \st\ is not a thing. Consequently, it cannot have
properties.}. This kind of pregeometry should be the consequence of
the unavoidable merging of Quantum Mechanics and General Relativity at 
very small distances. We emphasize that the pregeometry we propose
here is valid only for lengths above a minimum length, which is suggested to
be the Planck length by arguments based on the (yet unknown) 
theory of Quantum Gravity (Garay 1995).
\\

The structure of the paper is as follows: in Section 2 we offer a
brief account of the main ontological assumptions of the \th. Section
3 contains some formal tools, such as uniform spaces, to be used
later. The axiomatic core is presented in Section 4. Finally, in
Section 5 we give a short sketch of an extension of these ideas to Riemannian
space-times, we compare our theory with the theory of Bunge (1977) and, to
close, some observations on the nature of \st\ are pointed out.

\section{Ontological Background}

\hs In this section we give a brief synopsis of the ontological
presuppositions that we take for granted in our \th. For greater
detail see Bunge (1977, 1979) and Perez-Bergliaffa \et\ (1996). The basic 
statements of the ontology can be formulated as follows:

\begin{enumerate}

\item There exist concrete objects $x$, named {\em  things.} The set of
all the things is denoted by $\Theta$.

\item Things can juxtapose ($\dot{+}$) and superimpose ($\dot{\times}$) to 
give new things:
$$ 
x \dot{+} y = \{x,y\}\in \Theta  
$$ 
$$
x \dot{\times} y \in \Theta  
$$

\item The null thing is a fiction introduced in order to give a
structure of Boolean algebra to the laws of composition of things:
\begin{eqnarray*}
x \dot{+} \Diamond & = & x\\
x \dot{\times} \Diamond & = & \Diamond
\end{eqnarray*}

\item Two things are separated if they do not superimpose:
\[
x \wr y \Leftrightarrow x \dot{\times} y = \Diamond
\]

\item Let $T$ a set of things. The {\em aggregation} of $T$ (denoted $[T]$)
is the supremum of $T$ with respect to the operation $\dot{+}$.

\item The world $(\Box)$ is the aggregation of all things: 
\[ 
\Box = [\Theta] \Leftrightarrow (x \sqsubset \Box \Leftrightarrow x \in
	\Theta) 
\] 
where the symbol `$\sqsubset$' means `to be part of'. It stands for a 
relation between concrete things and should be not mistaken with `$\in$', which 
is a relation between elements and sets ({\em i.e.} abstract entities).

\item All things are made out of basic things $x\in {\Xi}\subset
\Theta$  by means of juxtaposition or superimposition. The basic
things are elementary or primitive:
\[ 
(x,y \in {\Xi}) \wedge (x \sqsubset y) \Rightarrow x=y 
\]

\item Things $x$ have {\em properties} $P(x)$. These properties can be
intrinsic or relational.


\item The {\em state} of a thing $x$ is a set of functions from a 
domain of reference $M$ to the set of properties ${\cal P}$.  The set
of the accesible states of a thing $x$ is the {\em lawful state space} of
$x$: $S_{\rm L}(x)$. The state of a thing is represented by a point in
$S_{\rm L}(x)$.

\item A {\em legal statement} is a restriction  upon the state
functions of a given class of things. A {\em natural law}
is a property represented by an empirically corroborated legal statement.

\item The {\em ontological history} $h(x)$ of a thing $x$ is a part of
$S_{\rm L}(x)$ defined by
\[
	h(x) = \{ \langle t, F(t) \rangle | t \in M\} 
\] 
where $t$ is an	element of some auxiliary set $M$,
and $F$ are the functions that represent the properties 
of $x$.

\item Two things {\em interact} if each of them modifies the history of the
other:
\[
x\Join y\Leftrightarrow h(x\dot{+}y)\neq h(x)\cup h(y)
\]

\item A thing  \(x_{\rm f}\) is a {\em reference frame} for \(x\) iff \\
(i) $M$ equals the state space of \(x_{\rm f}\), and\\
(ii) $ h(x  \dot{+} f) = h(x) \cup h(f)$

\item A {\em change} of a thing $x$ is an ordered pair of states:
\[ (s_1, s_2 ) \in E_{\rm L}(x) = S_{\rm L}(x) \times S_{\rm L}(x) \]

A change is called an {\em event}, and the space $E_{\rm L}(x)$ is called
the {\em event space} of $x$.

\item An event $e_1$ {\em precedes} another event $e_2$ if they
compose to give $e_3 \in E_{\rm L}(x)$:
\[
e_1 = ( s_1,s_2 )\; \land\;  e_2 = ( s_2,s_3 )
\Rightarrow  e_3 =  (s_1,s_3 )
\]
\end{enumerate}

The ontology sketched here (due mainly to M. Bunge) is realistic,
because it assumes the existence of things endowed with properties,
and objective, because it is free of any reference to cognoscent
subjects.\\

We will base the axiomatic formulation of the pregeometry of \st\ on
this ontology and on the formal tools that will be described in the
following.

\section{Formal Tools}

\subsection{Topological Spaces}

We give here just a brief review; for details the reader is referred to
Thron (1966) and references therein.

\begin{dfn}
${\cal P}(A)=_{\rm Df} \{X/X\subset A\}$ is the {\em power set} of the set $A$.
\end{dfn} 
\begin{dfn}
Let $A$ be a set. A subset ${\cal Z}$ of ${\cal P}(A)$ is a 
{\em topology} on $A$ if
\begin{enumerate}
\item $\emptyset\in {{\cal Z}}$, $A \in {\cal Z}$.

\item if $A_i \in {\cal Z}$, $i\in [i_1,...,i_n]$, then $\bigcup_{i=1}^n
A_i\in {\cal Z}$

\item if $A_i \in {\cal Z}$, $i\in [i_1,...,i_n]$, then $\bigcap_{i=1}^n
A_i\in {\cal Z}$
\end{enumerate}

The elements of ${{\cal Z}}$ are usually known as the {\em open sets}
of $A$. The pair $(A, {\cal Z})$ is called a {\em topological
space}. The elements of $A$ on which a topology ${\cal Z}$ is
defined are the {\em points} of the space $(A, {\cal Z})$
\end{dfn}

\begin{dfn}
A family ${\cal B}\in {\cal P}(A)$ is a {\em base} iff the family 
${\cal Z}$ of all unions of elements of ${\cal B}$ is a topology on
$\bigcup \{B/ B\in {\cal B}\}$. It is said then that 
${\cal Z}$ is the topology {\em generated} 
by ${\cal B}$.

\end{dfn}

\subsection{Filters}

\begin{dfn}
A nonempty family ${\cal B}$ of subsets of a set $A$ is a {\em filter}
on $A$ iff:
\begin{description}
\begin{enumerate}
\item  \(
A_1\in {\cal F} \wedge A_2 \in {\cal F}\Rightarrow A_1\cap A_2\in {\cal F}
\)
\item $B\supset A\in {\cal F}\Rightarrow B\in {\cal F}$ 

\item  $\emptyset\not\in {\cal F}$

\end{enumerate}
\end{description}
\end{dfn}

\begin{dfn}
A nonempty family $B$ of subsets of a set $A$ is called a 
{\em filter base} on $A$ provided ${\cal B}$ does not contain the
empty set and provided the intersection of any two elements of ${\cal
B}$ contains an element of ${\cal B}$
\end{dfn}
\subsection{Uniform spaces}

\begin{dfn}
 A nonvoid family $\Lambda$ of subsets of $A\times A$ is a
{\em uniformity} on $A$ iff

\begin{enumerate}

\item $L\supset\Delta = \{(x,x)/x\in A\}$ for all $L\in \Lambda$

\item $C\supset L\in \Lambda$ implies $C\in\Lambda$

\item If $L_1,L_2\in\Lambda\Rightarrow L_1\cap L_2\in\Lambda$

\item If $L\in\Lambda\Rightarrow L^{-1} = \{(x,y)/(y,x)\in L\}\in \Lambda$ 

\item For all $L\in \Lambda$ there exists a $K\in \Lambda$ such that
$K\circ K\subset L$, where $K\circ K = \{(x,y)/\exists z/(x,z)\in K,
(z,y)\in K\}$

\end{enumerate}
\end{dfn}
\begin{dfn}
The pair 
formed by $(A,\Lambda)$ is called a {\em uniform space}.
\end{dfn}
\begin{dfn}
 If $\Lambda$ satisfies that $\bigcap L\in\Lambda = \Delta\Rightarrow
\Lambda$ is a {\em separated (or Hausdorff) uniformity}. 

\end{dfn}
\begin{remark}
 Notice that a uniformity is a filter on $A\times A$ each
element of which contains $\Delta$. Property 4 is a symmetry property, 
whereas property 5 is an abstract version of the triangle inequality.
\end{remark}

\begin{dfn}
A set $B$ is called {\em everywhere dense} in a set A iff ${\bar B}$ (the
closure of $B$) $\supset A$.
\end{dfn}
\begin{dfn}
A topological space $(X, \tau )$ is {\em separable} iff there
exists an everywhere-dense subset of $X$ which is denumerable.
\end{dfn}

\begin{dfn}
 A filter ${\cal F}$ in a uniform space $(A, \Lambda)$ is called a 
{\em Cauchy filter} iff, given $L\in\Lambda$ there exist $M\in {\cal F}$ such 
that $M\times M\subset L$. A uniform space is {\em complete} 
iff every Cauchy filter has a limiting point.
\end{dfn}

\begin{thm}
Every separated uniform space has a {\em completion}. That is to say, 
one can always add ``ideal elements'' to complete the space. 
\end{thm}

\begin{proof} 
see Thron (1966), pp. 184-185. 
\end{proof}
 
\subsection{Metric spaces}
\label{MetricSpaces}

\begin{dfn}
Let $X$ be a set. A function $d:X \times X \mapsto \Re^+$ is a {\em metric} 
on $X$ iff:
\begin{enumerate}
\item $d(x,y) = 0 \Rightarrow x=y$ for all $x,y\in X$
\item If $x=y\Rightarrow d(x,y) = 0$ for all $x,y\in X$
\item $d(x,y) = d(y,x)$ for all $x,y\in X$
\item $d(x,y) + d(y,z) \geq d(x,z)$ for all $x,y,z\in X$     \label{M-Triang}
\end{enumerate}
\label{metricdef}
\end{dfn}

\begin{dfn}
The pair $(X,d)$ is a {\em metric space}. 
\end{dfn}
\begin{thm}
{\bf Theorem of metrization:}
A uniform space is {\em metrizable}
if and only if it is separable and its uniformity has a
numerable base (Kelley 1962).
\label{metriz}
\end{thm}
\begin{thm}
\label{isomcom}
{\bf Theorem of isometric completion:}
Any metric space is
isometric to a subspace dense in a complete metric space (Kelley 1962). 
\end{thm}

\begin{thm}
Let $S$ be a subset of $X$. Let $(X, {\cal H})$ a uniform space. Then the family 
${\cal H}_S = \{ H\bigcup(S\times S)/H \in {\cal H}$ is a uniformity on $S$
(called the relativized uniformity), and $\tau _{{\cal H}_S} = (\tau _{\cal H})
_S$.
\label{Thron175}
\end{thm} 

\section{Axiomatics}

We present now the axiomatic core of our formulation. The generating basis of 
primitive concepts is:\\

$B=\{\Xi, {\cal P}, S_{\rm L}, E_{\rm o}, E_{\rm G}, T_{\rm u},
 \dot{+}, \dot{\times}, \le, c\}.$\\

The different symbols are characterized by the ontological background (Sect. 2) 
and a set of specific axioms. We shall classify these 
axioms in ontological (o), formal (f), and semantical (s), according to
their status in the theory.

\begin{axm}
{\bf (o)} For each  $ x \in \Xi $ there exists a
single ordering relation:
\[ 
s_1 \le s_2 \iff s_2 = g(s_1) 
\] 
where $ g: S_{\rm L} \to S_{\rm L} $ is a legal statement.
\end{axm}

\begin{axm}
{\bf (s)} The set of legal states of $x$, $S_{\rm L}(x)$, is {\em temporally ordered} 
by the relation $\le$.
\end{axm}

\begin{dfn}
$s_1 \le s_2 \iff s_1$ {\em  precedes temporally } $s_2$.
\end{dfn}

\begin{remark}
The relation $\leq$ is a partially ordering relation:
there are states that are not ordered by $\leq$ (\eg\ given the initial
conditions $x_0$, $v_0$, there are states, which are characterized by the
values of $x$ and $v$, that can not be reached by a classical particle).
\end{remark}

\begin{dfn}
A subset of $S_{\rm L}(x)$ totally ordered by the relation $\leq$ is 
called {\em proper history of $x$}.
\end{dfn}
\begin{axm}
\label{HistUnique}
{\bf (o)} 				
For each thing $x$, there exists one and only one proper history.
\end{axm}

\begin{axm}
\label{OntCont}
{\bf (o)}				
If the entire set of states of an ontological history is
divided in two subsets $h_{\rm p}$ and $h_{\rm f}$ such that every
state in $h_{\rm p}$ temporally precedes any state in $h_{\rm f}$,
then there exists one and only one state $s_0$ such that $s_1\leq
s_0\leq s_2$, where $s_1\in h_{\rm p}$ and $ s_2\in h_{\rm f}$.  In
symbols:
\[
(\forall s_1)_{h_{\rm p}}(\forall s_2)_{h_{\rm f}} (s_1\leq s_2)(\exists
s_0)(s_1\leq s_0\leq s_2)
\]
\end{axm}



\begin{remark}
 This axiom expresses the notion of {\em ontological continuity}.
\end{remark}

\begin{dfn}
$h_{\rm p}$ is called the {\em past} of $s_0$, and $h_{\rm f}$ is called
the {\em future} of $s_0$.
\end{dfn}

\begin{remark}
Notice that past and future are meaningful concepts 
just when they are referred to a given state $s_0$.
\end{remark}

\begin{axm}
{\bf (o)} For every thing $x$, there exists another thing $x_t$ 
called {\em clock}, and an injective application $\psi$ such that:

\begin{enumerate}

\item $\psi_t :S_{\rm L}(x_t)\rightarrow S_{\rm L}(x)$

\item Given $t$, $t'\in S_{\rm L}(x_t)$: $t\leq t'\Rightarrow
\psi (t) \leq \psi (t')$
\end{enumerate}
\label{inj}
\end{axm}

\begin{thm} 				\label{T_1}
Given a thing $x$ with ontological history $h(x)$ and an
arbitrary system of units $U_{{\tau}}$ there exists a bijection:
\[{{\cal T}} : h\times U_{{\tau}} \leftrightarrow \Re \] 
that gives a parametrization $s_x=s_x({{\tau}})$.
\end{thm}

\begin{proof} 
From {\bf A\ref{HistUnique}}, {\bf A\ref{OntCont}}, and Rey Pastor \et\ (1952).
\end{proof}

\begin{dfn} 
\label{propti}
The variable ${{\tau}}$ is called {\em proper time} of $x$.
\end{dfn}

\begin{thm}				\label{T_2}
Let $x_t$ be a clock for $x$, with event space $E_{\rm L}(x_t)$, and $U_t$
an arbitrary system of units. There exists a bijection 
\[ 
T : E_{\rm L}(x_t)\times U_t \leftrightarrow \Re \] 
that provides a parametrization $s=s({{\tau}})$.
\end{thm}

\begin{proof} 
Generalization of {\bf T\ref{T_1}}. 
\end{proof}
\begin{dfn} 
$t$ is the {\em duration} of an event of $x$ respect to
the clock $x_t$.
\end{dfn}
 
This is what we need to say about time. For more details, see Bunge (1977).

\begin{axm}
\[
(\forall x)(x\sqsubset \Box)(\exists y)(y \sqsubset \Box \land y \Join x)
\]
\end{axm}
\begin{remark}
This latter axiom states that there exist no completely isolated things.
\end{remark}
We shall show now that the relation of interaction $\Join$
(see Sect. 2), generalized in a convenient way, induces a uniform 
structure (see Sect. 3) on the set of basic
things. It is important to note that the relation $\Join$ is symmetric
but neither reflexive nor transitive. However, it is always possible
to define a reflexive-transitive closure of a given relation (see
Salomaa 1973).  The closure $\Join^*$ of the relation of interaction
is the set of pairs of basic things that interact either directly or by
means of a chain (finite or infinte) of basic things.

Now the following theorem can be proved:

\begin{thm}				\label{T_4}
The relation $\Join ^*$ defines a uniform structure on ${\Xi}$.
\end{thm}

\begin{proof} 
Every equivalence relation defines a uniform structure on a
set (Thron 1966).
\end{proof}


\begin{remark} 
Armed with this theorem, we will be able to give
space a uniform structure. 
\end{remark}

In order to introduce the concept of space
we shall use the notion of {\em reflex action} between two
things. Intuitively, a thing $x$ acts on another thing $y$ if the
presence of $x$ disturbs the history of $y$. Events in the real world
seem to happen in such a way that it takes some time for the
action of $x$ to propagate up to $y$. This fact can be used to
construct a relational \th\ of space {\em \'a la} Leibniz, that is, 
by taking space as
a set of equitemporal things. It is necessary then to define the
relation of simultaneity between states of things.\\

Let $x$ and $y$ be two things with histories $h(x_{{\tau}} )$ and
$h(y_{{\tau}} )$, respectively, and let us suppose that the action of $x$
on $y$ starts at ${{\tau}}_x^0$. The history of $y$ will be modified
starting from ${{\tau}} _y^0$.  The proper times are still not related
but we can introduce the reflex action to define the notion of
simultaneity. The action of $y$ on $x$, started at ${{\tau}} _y^0$,
will modify $x$ from ${{\tau}} _x^1$ on. The relation ``the action of
$x$ is reflected on $y$ and goes back to $x$'' is the reflex action.
Historically, G. Galilei (1945) introduced the reflection of a light
pulse on a mirror to measure the speed of light. With this relation we
will define the concept of simultaneity of events that happen on
different basic things (see also Landau \& Lifshitz 1967).\\

We have already seen in Sect. 2 that a thing $x$ acts upon a thing $y$ 
if the presence of $x$ modifies the history of $y$: 
\[
x \rhd y =_{\rm Df} h(y|x) \neq h(y)
\]
where $h(y|x)$ represents the history of $y$
in the presence of $x$. 

The {\em total action} of $x$ upon $y$ is 
\[
{\cal A}(x,y) = h(y|x) \cap \overline{h(y)},
\]
where the overline denotes the complement.

Let us define now the history of $x$ after ${{\tau}}^0_x$
as 
\[
h(x,{{\tau}}^0_x) = h(x)|_{{{\tau}}_x > {{\tau}}^0_x}
\]
and similar definitions for $h(y,{{\tau}}^0_y)$ and for the history of $y$
after ${{\tau}}^0_y$ in the presence of $x$ after ${{\tau}}^0_x$, denoted
here as $h(\langle y,{{\tau}}^0_y \rangle, \langle x,{{\tau}}^0_x\rangle)$.

The total action of $x$ after $\tau^0_x$ on $y$ after ${{\tau}}^0_x$ is
\[
{\cal A}(y, x^0) = h(y|x) \cap \overline{h(\langle y,{{\tau}}^0_y \rangle,
\langle x,\tau^0_x\rangle)} 
\]
In a similar way we define the action of $y$ on $x$ posterior
to $\tau^1_y$.

$\tau^0_y$ is the minimum value of the proper time of $y$ for which
the action of $x$ posterior a $\tau^0_x$ is felt. 
\[
{{\tau}}^0_y = \inf \{{{\tau}}_y| {\cal A}(y,x^0) \}
\]

This quantity always exists, because of ontological
continuity, enforced in {\bf A\ref{OntCont}}. 
 
Similarly, we define ${{\tau}}^1_x$:
\[
{{\tau}}^1_x = \inf \{{{\tau}}_x| {\cal A}(x,y^0) \}
\]
Finally we can introduce a relation between the three instants
involved in the reflex action. We will call ${\cal R}<{{\tau}}^0_x,
{{\tau}}^0_y, {{\tau}}^1_x>$ the relation given by the set of ordered
3-tuples and established by the previous equations.\\

Let us go back to the axiomatics. 

\begin{axm}
{\bf (o)}
\label{A_7}
 Given two different and separated basic things $x$ and $y$, there exists a
minimum positive bound for the interval $({{\tau}}_x^1-{{\tau}}_x^0)$, defined 
by ${\cal R}$.
\end{axm}

\begin{remark}
Hereafter we shall deal only with 3-tuples
$<{{\tau}}_x^0,{{\tau}}_y^0,{{\tau}}_x^1>$ that satisfy the minimum
condition.
\end{remark}

\begin{dfn}
\label{simuldef}
${{\tau}}_y^0$ is simultaneous with $\tau^{1/2}_x =_{\rm Df}
1/2 ({{\tau}}_x^0 + {{\tau}}_x^1)$.
\end{dfn}

\begin{thm}
${{\tau}}_x$ and ${{\tau}}_y$ can be {\em synchronized} by the 
simultaneity relation.
\label{synchro}
\end{thm}

\begin{proof} There exists a bijection between $\tau _x$ and $\tau _y$
because ${\cal R}^{-1}$, the inverse of ${\cal R}$, is well-defined.
\end{proof}


{\bf Comment:} As we know from General Relativity, the simultaneity
relation is transitive only in special reference frames called {\em
synchronous} (Landau and Lifshitz 1967). We then include the
following axiom:

\begin{axm}
 {\bf (f)}
Given a set of basic things $\{ x_1, x_2,...\}$, there exists an
assignation of proper times ${{\tau}}_1, {{\tau}}_2,...$ such that the
relation of simultaneity is transitive.
\end{axm}

\begin{thm}
The relation of simultaneity is an equivalence relation.
\end{thm}

\begin{proof} 
From {\bf T\ref{synchro}} and {\bf A\ref{A_7}}.
\end{proof}
\begin{remark}
We should mention that, because of {\bf T\ref{T_1}}
and {\bf D\ref{propti}}, the history of a given 
thing is parametrized by its proper time $\tau$. Then, the relation of 
simultaneity is defined not over things but over states of things.
\end{remark}
\begin{dfn}				\label{OnticSpaceDef}
The equivalence class of states defined 
by the relation of simultaneity on the set of all basic things 
is the {\em ontic space} $E_{\rm o}$.
\end{dfn}

\begin{thm}
The ontic space $E_{\rm o}$ has a uniform structure.
\end{thm}

\begin{proof} 
Let $S$ be a set of states of things related by the simultaneity relation.
Because of the uniqueness of the ontological history 
postulated in {\bf A\ref{HistUnique}}, there is a one-to-one relation 
between a state in $S$ and a given thing, and then $S$ is 
isomorphic to a subset of $\Xi$. Then, by {\bf T\ref{Thron175}}, $S$ is a 
uniform space.
\end{proof}

\begin{axm}
 There exists a subset $D$ in the set of simultaneous states of 
interacting things $S$ that is denumerable and dense in $S$.
\end{axm}

\begin{remark}
This axiom requires space to be a {\em plenum}. Indeed, this hypothesis
(introduced by Aristotle and later supported by Leibniz) is 
central to Quantum Physics, and it permits the prediction of 
a plethora of vacuum phenomena (like the Casimir effect),
in good agreement with observation.  
\end{remark}

\begin{axm}
 Each $x\in \Xi$ interacts with a denumerable
set of basic things.
\label{denum}
\end{axm}

\begin{thm} 
The power set of $\Xi$ reduced to the equivalence
class is a basis for the uniformity (Bourbaki 1964).
\end{thm}

So now we are in the conditions of the th. of metrization:

\begin{thm}
The ontic space is metrizable.
\end{thm}

\begin{proof}
Immediate, from {\bf T\ref{metriz}}.
\end{proof}

\begin{dfn}
$\tau_u$ is the proper time of a reference thing $x_{\rm f}$ .
\end{dfn}

\begin{thm}
$(\forall x)_\Theta (\exists f_x) ( \tau_x=f_x^{-1}(\tau_u))$
\end{thm}
\begin{proof}
Immediate, from {\bf A\ref{inj}}.
\end{proof}
\begin{remark}
We shall call $\tau_u$ the {\em universal time}.
\end{remark}

The ontic space $E_{\rm o}$ is still devoid of any geometric
properties and consequently cannot represent the physical space. We
postulate then:

\begin{axm}
\label{LandauDist}
{\bf (f)} 				
The metrization of the ontic space is given by 
$$ d(x,y) = \frac{1}{2}c|\tau^1_{\rm x} -\tau ^0_{\rm x}| $$
where $c$ is a constant with appropriate dimensions, and the distance is
evaluated at $\tau^0_y$, which is simultaneous with $\tau^{1/2}_x$.
\end{axm}

\begin{thm}				\label{IsomCompOntic}
The ontic space is isometric to a subspace dense in a complete space.
\end{thm}

\begin{proof} 
The proof follows immediately form the theorem of 
isometric completion ({\bf T\ref{isomcom}}).
\end{proof}

\begin{dfn}				\label{GeomSpaceDef}
The complete space mentioned in ${\bf T\ref{IsomCompOntic}}$
is called  {\em geometric space} $E_{\rm G}$.
\end{dfn}
\begin{remark}
Because of the isometry mentioned in {\bf T\ref{IsomCompOntic}}, $E_{\rm G}$ 
inherits the metric of $E_{\rm o}$. Besides, note that every filter of Cauchy 
has a limiting point in $E_{\rm G}$, because this space is complete.
\end{remark}

\begin{dfn}
The elements of the completion are called {\em ideal things}.
\end{dfn}

\begin{remark}
 It should be noted that the ideal things (which are
abstract objects) do not belong to the ontic space but to
the geometric space.
\end{remark}

\begin{axm}
\label{GeomCond}
{\bf (f)} The points in $E_{\rm G}$ satisfy the following conditions:
\begin{enumerate}
\item Given two points $x$ and $y$ there exists a third point $y$ 
aligned with $x$ and $z$.
						\label{GeomCond1}

\item There exist three nonaligned points

\item There exist four noncoplanar points.

\item There exist only three spatial dimensions \footnote{This may not be 
true at the Planck scale. See for instance Tegmark 1997.}.

\end{enumerate}
\end{axm}

\begin{remark}
All the conditions in {\bf A\ref{GeomCond}} can be expressed in
terms of the distance \(d(a,b)\). For instance, 
(\ref{GeomCond1}) can be written in the form:
	\begin{eqnarray}
	(\forall(a))_G (\forall(b))_G (\exists c)_G  
		\left( d(a,b) + d(b,c)\right. & = & d(a,c)  \lor\\
		 d(a,c) + d(c,b)& = &d(a,b)  \lor\\
		 d(c,a) + d(a,b) & = &\left. d(a,c) \right) 
	\end{eqnarray}
	For details, see Blumenthal (1965).
\end{remark}
\begin{thm}
The geometric space $E_{\rm G}$ is globally Euclidean.
\end{thm}

\begin{proof} 
From {\bf A\ref{GeomCond}}, see Blumenthal (1965). 
\end{proof}


\begin{remark}
The ontic space $E_{\rm o}$ is not Euclidean but dense on an Euclidean
space (\ie\ $E_{\rm G}$). This is a consequence of the fact that a
{\em ``sequence of Cauchy of things''} does not have in general a thing as a
limit.
\end{remark}

\begin{dfn}
$T_{\rm u}$ is the the bijection alluded to in {\bf T\ref{T_2}} for the 
reference
thing $x_{\rm f}$ with proper time $\tau_u$.
\end{dfn}
\begin{thm}
There exists a nontrivial geometric structure on $E_{\rm G}\times T_{\rm u}$.
\end{thm}

\begin{proof}
Let us introduce a Cartesian coordinate system in $E_{\rm G}$, 
with origin located in the reference thing $x_{\rm f}$. From {\bf T\ref{T_2}}, 
{\bf D\ref{simuldef}} and {\bf A\ref{LandauDist}},
\begin{equation}
(t_{\rm y} -t_{\rm x})^2 = \left[\frac{d(y,x)}{c}\right]^2 
\label{cone}
\end{equation}
\end{proof}


This equation describes a sphere of radius $c\delta t$ centered at $x$. Then
the family of spheres $S(x, t_{\rm x})\subset E_{\rm G}\times T_{\rm u}$ defines
a geometric structure on $E_{\rm G}\times T_{\rm u}$.

\begin{axm}
 {\bf (o)}
The {\rm cones of action} determined by (\ref{cone}) are
independent of the reference thing $x_{\rm f}$.
\end{axm}

\begin{thm} 
 The quadratic form 
\begin{equation}
\delta s^2=(c\delta t)^2-(\delta \vec{r})^2 
\label{inter}
\end{equation}
is invariant under changes of the reference thing.
\end{thm}

\begin{cor}
 $E_{\rm G}\times T_{\rm u}$ has a Minkowskian structure.
\end{cor}

\begin{thm}
The only coordinate transformations that leave invariant
the quad\-rat\-ic form (\ref{inter}) are the Lorentz transformations.
\end{thm}

\begin{axm}
{\bf (s)}
$E_{\rm G}\times T_{\rm u}$ represents the physical space-time.
\end{axm}

This axiom completes the formulation of the theory. It should be noted that 
the spatial relations that we perceive are defined between
macroscopic (\ie\ composed)
things. Our system of axioms can handle also this situation (which strictly 
does not belong to Protophysics but to Physics), if we incorporate a 
specific model for
a given thing, which should be based on an explicit form of the interaction
between basic things.

\section{Further comments}

\subsection{Extension to quantum basic things}

\hs The relational theory of space-time that 
we developed in the precedent section 
is based on the concept of basic thing. We should remark that this is 
a theory-dependent concept, \ie\ two different theories may take different 
sets of things as basic. In this sense, our theory 
is classical (as opposed to quantum), because {\bf T\ref{metriz}} enforces
the separability of basic things. However, it can be conveniently
modified to serve as a part of the protophysics of Quantum Theories, and
this we shall do in the following.\\

The main problem is the fact that quantum particles can superimpose (\ie\
the distance between two of them can be zero while the particles are
still distinguishable).
To incorporate this fact, we shall relax {\bf D\ref{metricdef}}, keeping only
items (2), (3), and (4). With this modification, 
{\bf D\ref{metricdef}} defines a pseudometric instead of a metric (Kelley 1962).
\\

Now we replace {\bf T\ref{metriz}} with
\begin{thm}
{\bf Theorem of pseudometrization:} a uniform space is pseudometrizable
if its uniformity has a denumerable base.
\end{thm}

So, due to {\bf A\ref{denum}}, $E_{\rm o}$ is pseudometrizable. Morover, it can be 
completed:
\begin{thm}
Every pseudometric space is isometric to a subspace dense in a complete
pseudometric space.
\end{thm}
We shall call this last space $E_p$ ({\em pregeometric space}). 
But we know that 
in Quantum Mechanics the Euclidean space is included in the corresponding 
protophysics (Perez Bergliaffa \et\ 1993). To recover Euclidean space, 
we begin by introducing the concept of ontic point:

\begin{dfn}
Let $X\sqsubset\Xi$ be a family of basic things. We say that $X$ is a 
{\em complete family of partially superimposed things} if
\begin{enumerate}
\item $(\forall x)_X (\forall y)_X (x\dot{\times}y \neq \diamond)$
\item $(\forall x)_{\bar{X}} (\exists y)_X (x\wr y)$
\end{enumerate}
\end{dfn}

We shall call this kind of family an {\em ontic point}, because the 
(pseudometric) distance between any two components of the family is zero.

\begin{axm}
Let $\xi$ and $\eta$ two ontic points. Then
$(\forall x_i)_\xi (\forall y_j)_\eta (\exists\; C(\xi, \eta))(d_p(x_i,y_j)<
C(\xi ,\eta))$.
\label{exist}
\end{axm}

\begin{dfn}
Let $\xi$ and $\eta$ be two ontic points.
The distance between them is given by
\begin{equation}
d_G (\xi ,\eta) =_{\rm Df} {\rm sup}_{(i,j)} d_p (x_i,y_j)
\label{disfun}
\end {equation}
with $x_i\in\xi$ and $y_j\in \eta$.
\end{dfn}

\begin{remark}
The axiom {\bf A\ref{exist}} guarantees that this distance is well-defined.
\end{remark}

\begin{thm}
The set of ontic points, toghether with the distance function (\ref{disfun})
is a metric space.\\

{\bf Proof:} the items (2), (3), and (4) of {\bf D\ref{metricdef}} are trivially 
satisfied because $d_p$ is a pseudometric. Regarding the first item, if $\xi$
and $\eta$ are two ontic points, there exist $x_i\in\xi$  and $y_j\in\eta$
such that $x_i\wr y_j$. The condition $d_p(x_i,y_j)>0$ is satisfied because
$d_p$ is a pseudometric. Then, $\xi\neq\eta\Rightarrow d_G(\xi ,\eta)>0$, 
which can be written as $d_G(\xi , \eta)=0\Rightarrow \xi =\eta$.
\end{thm}

\begin{thm}
{\bf Theorem of isometric completion:} any metric space is isometric to
a subspace dense in a complete metric space (Kelley 1962).
\end{thm}

\begin{dfn}
The completation of the space of ontic points is the geometric space $E_{\rm G}$.
\end{dfn}

From this point, the construction goes on as in the previous case.

\subsection{Extension to Riemannian spaces}

\hs Our theory is a pregeometry for a Minkowskian
space-time. Gravitational physics, however, requieres more complex
structures. In this section we shall
sketch the necessary steps that lead to a Riemanian space, which is
used in General Relativity (Covarrubias, 1993), in an informal
way, avoiding the technicalities. \\

A Riemannian space can be obtained from our theory using
a tetrad formulation. For this we need at least an axiom
elucidating the connection between ontic and geometric spaces:

\begin{axm}
{\bf (f)} The geometric space $E_{\rm G}$ is the tangent space to the ontic
space at the reference thing $x_{\rm f}$.
\end{axm}

With this axiom, the connection between ontic and geometrical
spaces will be purely local. The full space will be constructed by
pasting together patches of quasieclidean pieces. The following axiom
sketches the way this can be done:

\begin{axm}
 {\bf (f)}
	There exists a parallel displacement operator \(\omega\)
	connecting the components of vectors (\ie\ elements of the
	tangent space) tangent to $E_{\rm o}$ on neighbouring things.
\end{axm}

With the parallel displacement, a covariant derivative can be
defined in the usual way. The usual property of the Riemannian
conection (Ricci coefficients) must be posited. The following axiom
will do the job:

\begin{axm}
 {\bf (f)}
The covariant derivative $\nabla$ annihilates the metric.
\end{axm}

Since we are working
with a transitive simultaneity relation (which is equivalent to use a
synchronous reference frame in any metric theory of gravitation (Landau
\& Lifchitz 1967) a Riemannian space will define a unique pseudoriemannian
spacetime. In this way a protophysics for a rigurous formulation of
General Relativity (such as Covarrubias (1993)) and more general
theories of gravitation will be obtained. \\

The above scheme must be completed in several ways. Accurate
definitions should be given of the different constructs defined. Also
several axioms should be introduced to ensure a differential manifold
structure on a suitable completition of the ontic space. We shall not
pursue further this matter here, but in a future communication.

\subsection{Comparison with the theory of Bunge}

\hs As we have mentioned in the Introduction, protophysical
theories can be classified as subjective and objective, according to
whether cognoscent subjects and/or sensorial fields are considered as
basic objects or not. Bunge (1977) developed an
objective and realistic relational theory of space time 
and made a clear cut comparison with other subjective and objective
theories. In this section, we shall limit
ourselves to a comparison of the objective and realistic theory of
Bunge (1977) with the one developed in the present paper. \\
	
	Bunge's theory of space is based on the {\em
interposition relation} \((x|y|z)\) that can be read ``$y$ interposes
between $x$ and $z$''. The properties of this relation are posited in
Axioms 6.1 to 6.6 of Bunge (1977). In this section we shall
show how the corresponding relation can be constructed in the present
theory. \\

We shall first define a similar relation between basic things.

\begin{dfn}
	Let \(x, y, z \in \Xi\). We shall say that $[ x|y|z]_\Xi$
	if
	\[
 \left(d(x,y) + d(y,z) = d(x,z)\right) \land
	\left(x \wr y \wr z \wr x ) \lor (x=y=z)\right)
\]
\end{dfn}
	The next theorem proves that in our theory, the interposition
relation holds between basic things if and only if it is valid in 
Bunge's theory: 

\begin{thm}
 Let \(x, y, z \in \Xi\). Then $[ x|y|z]_\Xi$ if \( (x|y|z)\).
\end{thm}
\begin{proof}
The proof consists in showing that $[ x|y|z]_\Xi $ satisfies
	each of the seven conditions {\it (i)--(vii)} of Axiom 6.1 in
	Bunge (1977).
\end{proof}

	In order to define an interposition relation for general
things we shall use our interposition relation for basic things:

\begin{dfn}
Let \(\xi, \eta, \zeta \in \Theta\). Then $[\xi | \eta | \zeta]_\Theta$
either if they are equal (\(\xi=\eta=\zeta \)) or if there exists
three separate basic things \(x, y, z \in \Xi\) that are parts of
one thing but not of the others  and that interpose.
\begin{eqnarray}
[\xi | \eta | \zeta]_\Theta & =_{\rm Df}& (\xi=\eta=\zeta) \lor \nonumber\\
	& &\exists(x,y,z \in \Xi)\left\{
		\left[ (x \sqsubset \xi) \land (x \wr \eta) 
			\land (x \wr \zeta) \right] \land \right.
							\nonumber\\
	& &     \left[ (y \sqsubset \eta) \land (y \wr \xi) 
			\land (y \wr \zeta) \right]\land \nonumber\\
	& &     \left[ (z \sqsubset \zeta) \land (z \wr \eta) 
			\land (z \wr \xi) \right] \land \nonumber\\
	& &     \left. \left[x|y|z\right]_\Xi
					\right\}        
\end{eqnarray}
\end{dfn}
\begin{thm}
Let \(\xi, \eta, \zeta \in \Theta\).  Then \(\left[\xi|\eta|\zeta\right]_
\Theta\) if \((\xi|\eta|\zeta)\).
\end{thm}

In the same way, with appropriate definitions, it is possible
to show that the remaining postulates of Bunge's theory can be recovered as 
theorems
in our formulation.\\

The time theory exposed in the first part of our axiomatics is essentially the 
same
theory exposed in Bunge (1977), although our axioms are somewhat
different. The main differences lies in Axms. \ref{HistUnique} and
\ref{OntCont}. The first one, not explicitly stated in Bunge (1977),
forbids ``gardens of bifurcating paths'' (Borges, 1967) or, in
general, more than one time-like direction. 
The second axiom may be taken as a
reformulation of the heraclitean principle: ``Panta rhei''.\\

	From a formal point of view, the present theory of space-time is very
different from the theory in Bunge (1977). This is because our
fundamental relation of ``reciprocal action'' is very limitative and
the related axioms are extremely strong: we are led almost without
ambiguity to a Minkowskian structure of space-time. \\

	Finally, it is important to remark that, since both theories have 
the same referents
(namely, things and their properties), they are
referentially equivalent, realistic and objective relational
theories of space and time.

\subsection{The nature of \st}

	In the present theory, \st\ is not a thing but a substantial
property of the largest system of things, the world \(\Box\),
emerging from the set of the relational properties of basic
things. Thus, any existential quantification over \st\ can be
translated into quantification over basic things. This shows that \st\
has no ontological independence, but it is the product of the
interrelation between basic ontological building blocks. For instance,
rather than stating ``\st\ possesses a metric'', it should be said:
``the evolution of interacting things can be described attributing a
metric tensor to their spatio-temporal relationships''. In the present 
theory, however, \st\ is intrepreted
in an strictly materialistic and Leibnitian sense: it is an
order of succesive material coexistents.

\section{Summary}

We have developed a materialistic
relational theory of \st, that carries out the program initiated by 
Leibniz, and provides a 
protophysical basis consistent with any
rigorous formulation of General Relativity. Space-time is constructed 
from general concepts which are common to any consistent scientific 
theory. It is shown, consequently, that there is no need for positing the 
independent existence of space-time over the set of individual things.  

\section{Acknowledgements}

The authors are indebted, in many different ways, to M. Bunge,
O. Barraza, M. Castagnino, A. Ord\'o\~nez, P. Sisterna, J. Horvath, 
F. Gaioli, and E. Garc\'\i{}a \'Alvarez  for discussions,
criticisms and help. They also would like to acknowledge economic support
from CONICET, FAPESP, UNLP, USP and ICTP during the long development of
this paper.

\end{document}